\documentclass{emulateapj}

\newcommand{\lya}{Ly$\alpha$}

\newcommand{\sub}[1]{\ensuremath{_\textrm{\scriptsize{#1}}}}

\shorttitle{Fluorescent Constraints on QSO Lifetimes}
\shortauthors{Trainor \& Steidel}

\begin{document}


\title{Constraints on Hyperluminous QSO Lifetimes via Fluorescent \lya\ Emitters at $Z\simeq2.7$}


\author{Ryan Trainor}
\author{Charles C. Steidel}
\affil{Cahill Center for Astrophysics, MC 249-17, 1200 E California Blvd, Pasadena, CA 91125}




\begin{abstract}
We present observations of a population of \lya\ emitters (LAEs) exhibiting fluorescent
emission via the reprocessing of ionizing radiation from nearby
hyperluminous QSOs. These LAEs are part of a survey at redshifts $2.5
< z < 2.9$ combining narrow-band photometric selection
and spectroscopic follow-up to characterize the emission mechanisms,
physical properties, and three-dimensional locations of the emitters
with respect to their nearby hyperluminous QSOs. These data allow us
to probe the radiation field, and thus the radiative history, of the
QSOs, and we determine that most of the 8 QSOs in our sample have been
active and of comparable luminosity for a time 1 Myr $\lesssim
t\sub{Q} \lesssim 20$ Myr. Furthermore, we find that the
ionizing QSO emission must have an opening angle $\theta\sim 30\degr$
or larger relative to the line of sight. 
\end{abstract}





\section{Introduction} \label{sec:intro}

The extreme luminosities of QSOs make them effective tracers of
black hole growth over most of the Universe's history ($0 < z \lesssim
7$) and likewise illuminate the evolution of galaxies and large-scale
structure over these cosmological epochs and volumes. In particular,
the mass of supermassive black holes is tightly
correlated with properties of their host galaxies
(e.g. \citealt{mag98,geb00,fer00}), while QSO clustering properties
are well-matched to the expected distribution of dark
matter in and out of halos (e.g. \citealt{sel00,ber02,zeh04})
according to the cosmological standard model.

However, the manner in which QSOs populate and interact with galaxies
and dark-matter halos is obscured by the unknown timescales and
geometries over which black holes accrete mass and produce substantial
radiation. Specifically, the fraction of black holes that can be
observed as QSOs depends sensitively on the length of
their active phases and whether their emission is isotropic or confined
to a narrow solid angle. A short QSO lifetime ($t\sub{Q}$) and/or a small opening angle
($\theta\sub{Q}$) would indicate that observed QSOs comprise a small
fraction of the total population of black holes (and therefore galaxies and
dark-matter halos) that will pass through a QSO phase. Furthermore, both the
comparison of $t\sub{Q}$ to a typical star-formation timescale
$t\sub{SF}\sim100$ Myr and the value of $\theta\sub{Q}$ have deep
implications for the mechanisms by which the QSO couples to the ISM in
galaxies and produces feedback.

Current estimates of $t\sub{Q}$ utilize numerous methods, many of
which are described in detail in a review by \citet{mar04}, and
in general allow for QSO lifetimes in the range $10^6$ yr $<t\sub{Q}< 10^8$
yr. In the last decade, measurements of the QSO luminosity function
and clustering have provided particularly powerful constraints on
a globally-averaged $t\sub{Q}$ (e.g. \citealt{kel10}), but
they also rely on the poorly-constrained black hole mass function or
assume that the most luminous QSOs populate the most massive halos,
contrary to some observations and physically-motivated models
(e.g. \citealt{tra12,ade05e,con13}). These global measures of $t\sub{Q}$
are also degenerate between single-phase QSO accretion (in which the
bulk of activity occurs in a single event) and multi-burst models (in
which the same total time in a QSO phase is distributed over many
short accretion events).

More direct measurements of $t\sub{Q}$ may be obtained from
the effect of QSO radiation on their local enviroments.
Measures of the transverse proximity effect
use the volumes (and associated light-travel times) over which
bright QSOs ionize their nearby gas and have yielded estimates or
lower limits in the range
$t\sub{Q} \sim 16-40$ Myr by tracing \ion{He}{2}
(e.g. \citealt{jak03,wor07}) and metal-line \citep{gon08} absorption
systems. 

The detection of fluorescent \lya\ emission provides another direct
measurement of $t\sub{Q}$. Fluorescent \lya\ arises from the
reprocessing of ionizing photons (either from the
metagalactic UV background or a local QSO) at the surfaces of dense,
neutral clouds of \ion{H}{1}. This effect has been modeled with
increasing complexity over the last few decades
\citep{hog87,gou96,can05,kol10} and has now been
observed around bright QSOs
(\citealt{ade06,can07,can12}; Trainor et al. 2013, in prep.). These
detections have verified the feasibility of identifying fluorescent
emission from
the intergalactic medium (IGM) and suggested a wide opening angle
$\theta\sub{Q}$ of QSO emission (but see also \citealt{hen13}). They
are consistent with other constraints on $t\sub{Q}$, but have
been limited by small sample sizes and/or a lack of the 3D spatial
information necessary to probe the spatial variation of the QSO
radiation field in detail. 

This paper presents results from a large survey of \lya\ emitters
(LAEs) in fields surrounding hyperluminous ($L\sub{UV}\sim 10^{14}$
L\sub{$\sun$}) QSOs at redshifts $2.5 < z < 2.9$. The full results of this
survey, including a detailed analysis of the \lya\ emission mechanisms
and physical properties of the LAEs, will be presented in Trainor et
al. (2013, in prep.; hereafter Paper II). This paper focuses on the
implications of these data for $t\sub{Q}$ and $\theta\sub{Q}$.
Observations and the data are briefly discussed in
\S\ref{sec:obs}; identification of fluorescent sources is discussed in
\S\ref{subsec:fluor}; constraints on $t\sub{Q}$ and $\theta\sub{Q}$
appear in \S\ref{subsec:tdelay}$-$\S\ref{subsec:theta}; and conclusions are given in
\S\ref{sec:conclusions}. A standard cosmology with $H_0=70$ km s$^{-1}$
and $\{\Omega\sub{m},\Omega_\Lambda\} = \{0.3,0.7\}$ is assumed
throughout, and distances are given in physical units (e.g. pMpc).



\section{Observations} \label{sec:obs}


We conducted deep imaging in each survey field using custom
narrow-band filters and corresponding broad-band filters sampling the
continuum near \lya\ with Keck 1/LRIS-B over several
years; these data are described in Paper II. All 8 fields are part of the
Keck Baryonic Structure Survey (KBSS; \citealt{rud12a}; Steidel et
al. 2013, in prep.)

We used 1 of 4 narrow-band filters to image each of the 8 QSO fields
(centered on or near the QSO); a brief
description of the fields and corresponding filters is given in Table
\ref{table:fields}. Each filter has a FWHM $\sim80$\AA\ and a
central wavelength tuned to \lya\ at the redshift of one or more of
the hyperluminous QSOs. The QSOs span a redshift range $2.573 \le z
\le 2.843$, and the filter width corresponds to $\Delta z \approx 0.066$
or $\Delta v \approx 5400$ km s$^{-1}$ at their median
redshift. The narrow-band images have total integration times of 5-7
hours and reach a depth for point sources of $m\sub{NB}(3\sigma)\sim 26.7$;
the continuum images are typically deeper by $\sim 1.5$ mag.

Object identification and narrow-band and continuum magnitude
measurements were performed with SExtractor. The success rate of our
initial follow-up spectroscopy dropped sharply above $m\sub{NB}=26.5$,
so our LAEs are selected to have
$20<m\sub{NB}<26.5$ and $m\sub{cont}-m\sub{NB}>0.6$ (corresponding to a
rest-frame equivalent width in \lya\ $W_0\gtrsim 20$\AA). These
criteria define a set of 841 LAEs. The LAEs range from unresolved/compact to
extremely extended (FWHM $\gtrsim 10\arcsec$) sources. The extended objects
require large photometric apertures that may enclose
unassociated continuum sources; in order to avoid the complication of
determining the true continuum counterparts of these sources, LAEs
with FWHM $>3\arcsec$ were removed from our sample for this analysis,
leaving a final photometric sample of 816 LAEs.

Spectra were obtained with Keck 1/LRIS-B in the multislit mode using the
600/4000 grism; the spectral resolution near \lya\
for these spectra is $R\sim 1100$. Redshifts were measured
via an automated algorithm described in Paper II;
the detected \lya\ lines have a minimum SNR $\sim$ 3.5 and a median SNR $\sim$
15. Detected lines were required to lie at a wavelength where the
transmission exceeds 10\% for the narrow-band filter used to select
their associated candidates (typically a $\sim$110\AA\ range), but no
other prior was used to constrain the  
automatic line detection. Redshifts were measured for 260 of the LAEs
meeting our final photometric criteria; these comprise the
spectroscopic sample used in this paper. QSO redshifts were determined
as described in \citet{tra12} and have estimated uncertainties
$\sigma\sub{z,QSO}\approx 270$ km s$^{-1}$.




\section{Results} \label{sec:results}
\subsection{Detection of Fluorescent Emission} \label{subsec:fluor}

\lya\ emission is subject to complex radiative-transfer effects and is
ubiquitous in star-forming galaxies, so accurate discrimination
between fluorescence and other \lya\ emission mechanisms is key to its
use as a QSO probe (see \citealt{can05,kol10}). A powerful tool in this
regard is the rest-frame equivalent width of \lya\ ($W_0$), which has a
natural maximum in star-forming sources because the same
massive stars produce the UV continuum and the ionizing
photons that are reprocessed as \lya\ emission.
Empirically, UV-continuum-selected galaxies at $2\lesssim z \lesssim
3$ almost never exhibit  $W_0 >100$\AA\ \citep{kor10}, which is also
the maximum value expected for continuous star-formation lasting
$\sim10^8$ yr or longer \citep{ste11}. This threshold may demarcate
the realm of fluorescence from that of typical star-forming galaxies.
Furthermore, models of star-formation in the extreme limits of
metallicity and short bursts predict a stringent limit of $W_0<240$
\AA\ (see \citealt{sch02} and discussion in \citealt{can12}) for even
the most atypical star-forming galaxies.

\begin{figure*}
\includegraphics[scale=0.8]{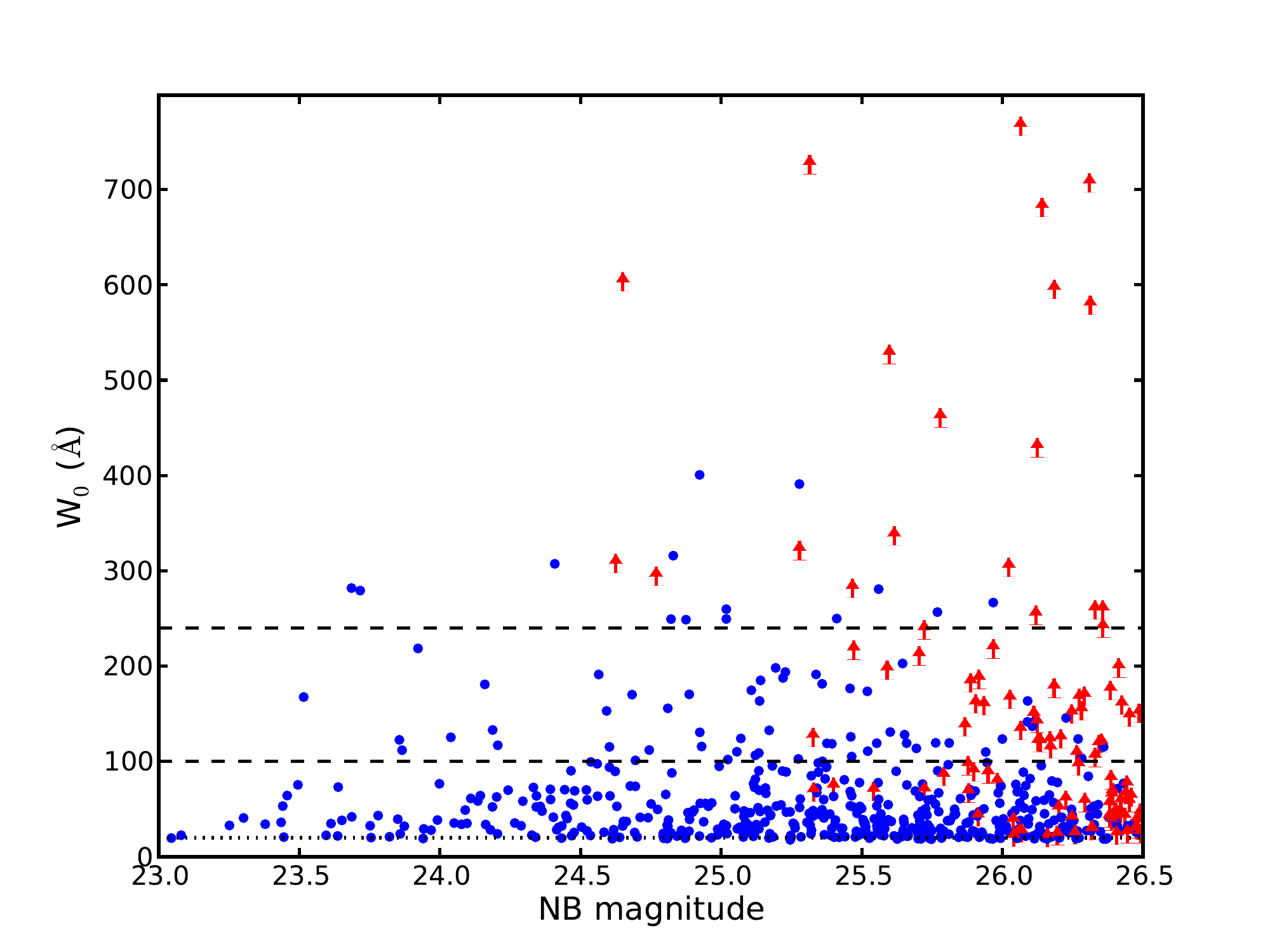}
\caption{The rest-frame equivalent width ($W_0$) of \lya\ determined
  from the measured $m\sub{cont}-m\sub{NB}$ color for each
  LAE in our photometric sample. Objects detected in the continuum
  band are displayed as blue
circles, while objects consistent with having no continuum flux (after
subtracting the measured \lya\ flux) are displayed as red arrows
denoting 1$\sigma$ lower limits. Thresholds for
the maximum value of $W_{0,Ly\alpha}$ consistent with typical and
extremely brief star-formation are
plotted as dashed lines at $W_0=100$\AA\ and $W_0=240$\AA,
respectively (see \S\ref{subsec:fluor} for details). The dotted line
at $W_0=20$\AA\ denotes the minimum \lya\ equivalent width for a LAE.}
\label{fig:nbma}
\end{figure*}

The distribution of $W_0$ and $m\sub{NB}$ for the LAEs are displayed
in Fig. \ref{fig:nbma}. Observed-frame equivalent widths ($W$) were
measured from the narrow-band color excess while accounting for the
presence of \lya\ within the continuum filter bandpass; details of
this procedure are given in Paper II. $W_0$ is estimated from $W$ by
$W_0=W/(1+z\sub{QSO})$, where $z\sub{QSO}$ is the redshift of the
associated hyperluminous QSO.

Fig. \ref{fig:nbma} demonstrates that many of our LAEs exceed both the
model and empirical thresholds for the maximum $W_0$ consistent with
pure star formation: of the 816 LAEs, 116 exceed $W_0>100$\AA\ and 32
exceed $W_0>240$\AA. We consider these high-$W_0$ sources as
excellent candidates to be fluorescence-dominated, with minimal 
stellar contribution to their \lya\ flux.

\subsection{Time delay of fluorescent emission}
\label{subsec:tdelay}

The presence of a fluorescent emitter indicates that its local
volume element was illuminated by ionizing QSO radiation at the
lookback time $t\sub{\lya}$ when the observed \lya\ photon was
emitted. Depending on whether the fluorescent source
lies in the foreground or background of the QSO, $t\sub{\lya}$ may be
greater or less than $t\sub{QSO}$, the lookback time to the QSO
itself.

However, this fluorescent \lya\ photon was generated by the
reprocessing of an ionizing QSO photon emitted a time $t\sub{lt}$
before $t\sub{\lya}$, where $t\sub{lt}$ is the light-travel time from
the QSO to the emitter. It is geometrically trivial to show that
$t\sub{QSO} \le t\sub{\lya}+t\sub{lt}$, and we can define the delay
time for an emitter at a vector position ${\bf r}$:

\begin{equation}
t\sub{delay}({\bf r}) = t\sub{\lya}({\bf r})+t\sub{lt}({\bf
  r})-t\sub{QSO} \,\, .
\label{eq:tdelay}
\end{equation}

It can likewise be shown that the locus of points for which
$t\sub{delay}$ is constant forms a paraboloid pointed toward the
observer with the QSO at the focal point (see Fig. \ref{fig:zcombo},
top, for a pictorial representation at varying values of
$t\sub{delay}$). 

The significance of such a delay surface can be seen by considering a
simple, step-function model of QSO emission in which the ionizing
emission was zero a time $t\sub{Q}$ before we observe the QSO and has
been constant since that time. Under such a model, the ionizing field
is zero for $t\sub{delay}({\bf r})>t\sub{Q}$, and the spatial
distribution of fluorescent emitters must be restricted to the
interior of the paraboloid defined by $t\sub{delay}({\bf r})=t\sub{Q}$.

Since the high-$W_0$ LAEs described in \S\ref{subsec:fluor} are highly
likely to be dominated by fluorescence, they represent the best
tracers of the QSO ionizing field. Below, we use the distribution of
these sources to set limits on $t\sub{Q}$.

\subsection{Constraints on $t\sub{Q}$ from the redshift distribution}
\label{subsec:zdist} 
\begin{figure*}
\includegraphics[scale=0.8]{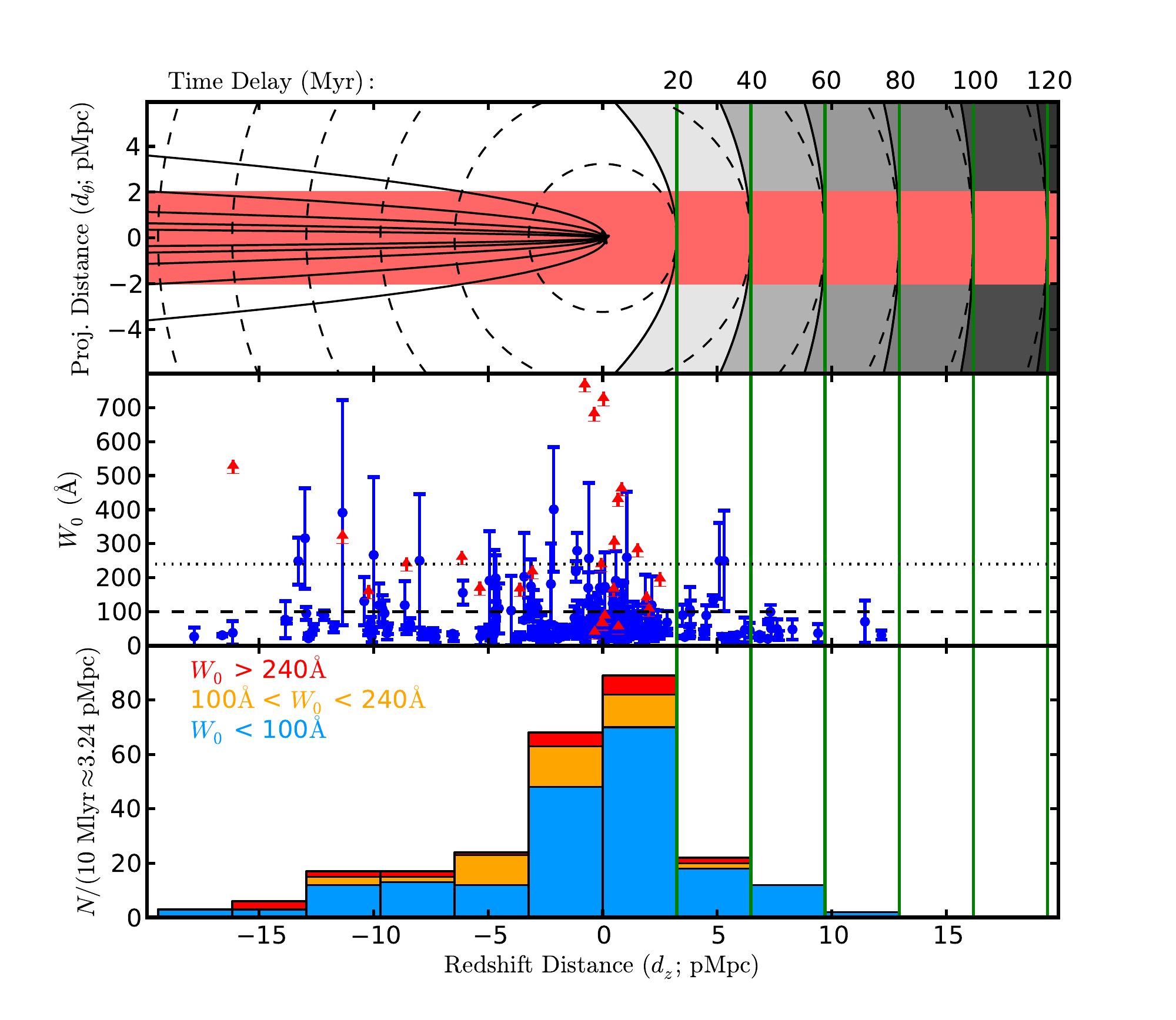}
\caption{Redshift constraints on QSO lifetimes. {\it Top:} Schematic
  representation of QSO light travel with observer lying to the left
  of the plot. Dashed circles represent the  current position of
  photons emitted by the QSO 10 Myr, 20 Myr, ... 60 Myr ago. Shaded
  parabolas on right side denote surfaces of constant $t\sub{delay}$. For
  the simple model of QSO emission discussed in \S\ref{subsec:qsos},
  sources may only exhibit fluorescent emission if they lie in the
  foreground of the surface for which
  $t\sub{delay}=t\sub{Q}$, where $t\sub{Q}$ is
  the QSO lifetime. The red region shows the geometry of this survey,
  for which surfaces of constant time delay are well-approximated by
  surfaces of constant redshift (green lines) for $t\sub{delay}\gtrsim
  20$ Myr. The narrow parabolas on
the left side are time-delay surfaces for $t\sub{delay}\le$ 1
Myr, which are shown in more detail in Fig. \ref{fig:rcombo}. {\it
  Middle:} \lya\ equivalent width ($W_0$) as a function of
line-of-sight distance from the QSO for LAEs with measured
redshifts. Red limits and blue points are plotted as in
Fig. \ref{fig:nbma} with the addition of 1$\sigma$ error bars. Nearly
all points with $W_0>100$\AA\ lie in the foreground of the
$t\sub{delay}=20$ Myr surface. {\it Bottom:} The redshift
distribution of emitters in three bins of $W_0$, where $W_0$ is either
the detected value or the 1$\sigma$ lower limit. The fraction of
emitters with $W_0>100$\AA\ drops sharply for $d\sub{z}>10$ Mlyr
$\approx$ 3.24 pMpc. }
\label{fig:zcombo}
\end{figure*}

The geometry of our survey volume is such that we can probe
long timescales $(10^7$ yr $\lesssim t\sub{Q} \lesssim 10^8$ yr$)$ via
the line-of-sight distribution of sources (see red box in
Fig. \ref{fig:zcombo}, top). The distribution of $W_0$ vs. $d\sub{z}$
for the 260 LAEs with measured redshifts is displayed in
Fig. \ref{fig:zcombo} (middle), where $d\sub{z}$ is the Hubble
distance of each emitter from its associated hyperluminous QSO
determined from the difference of emitter and QSO redshifts. 

Due to the effects of redshifts errors and peculiar velocities among
the LAEs and QSOs, our foreground/background discrimination breaks
down at $\Delta v \approx 700$ km s$^{-1}$, corresponding to
$d\sub{z}\lesssim 3$ pMpc and
$t\sub{delay}\lesssim 20$ Myr; we are therefore unable to constrain
$t\sub{Q}$ lower than 20 Myr based on the redshift distribution of
sources. At $t\sub{delay}>20$ Myr, however, there is a significant
paucity of high-$W_0$ sources: only four sources with $W_0>100$\AA\
lie in this range, and there are none at
$t\sub{delay}>33$ Myr. This suggests $t\sub{Q}\lesssim20$ Myr for
these fields; we evaluate the significance of this result via
numerical tests below.

The non-uniform redshift coverage of our QSO fields complicates the
analysis of $d\sub{z}$; the observed redshift distribution is
modulated by the narrow-band filter bandpass and the distribution of
large-scale structure along the QSO line of sight. In addition, the
QSO redshifts are not perfectly centered in their filter bandpasses,
which contributes part of the observed asymmetry. Fortunately, the
distribution of star-forming galaxies presumably represented by our
$W_0<100$\AA\ LAEs trace all of these effects, and we can use their
redshift distribution to determine the expected distribution of
emitters in the absence of fluorescence. We use the entire sample
of emitters with spectroscopic redshifts and $W_0<100$\AA\ for this
comparison.

Firstly, we utilize the two-sample Kolmogorov-Smirnov (KS) test, a
measure of the probability that two samples of observations are drawn from
the same parent distribution. Evaluating the test on
the distributions of $d\sub{z}(W_0>100$\AA$)$ and $d\sub{z}(W_0\le
100$\AA$)$ yields a KS statistic
$M\sub{KS}=0.234$, corresponding to $p<4\times 10^{-3}$. There are
insufficient emitters with redshift measurements to perform the test on
the higher-threshold ($W_0>240$\AA) sample with significance.

A disadvantage of the two-sample KS test is that it is most sensitive
to differences that occur near the center of two
distributions, whereas we expect the strongest deviation in these
distributions at large $d\sub{z}$, where no fluorescent emitters would
appear if $t\sub{Q}\lesssim 20$ Myr. For this reason, we 
conducted a Monte-Carlo test in which subsamples of the $W_0<100$\AA\
LAEs were randomly selected (with replacement), where each subsample
has 67 objects, the number of $W_0>100$\AA\ emitters with
spectroscopic redshifts in our actual sample. For each subsample, the
number of sources with $d\sub{z}>3$ pMpc (corresponding to
$t\sub{delay}\gtrsim 20$ Myr) was counted. 

As noted above, only four sources with $W_0>100$\AA\ exceed
$d\sub{z}=3$ pMpc. In $10^5$ simulated subsamples, 40 had three or
fewer objects with $d\sub{z}>3$ pMpc, 
yielding a significance $p<4\times 10^{-4}$. We therefore conclude
that the 7 QSOs with spectroscopic follow-up (ie. all except Q2343+12)
are consistent with $t\sub{Q}\lesssim 20$ Myr (approximately the
lowest it can be measured unambiguously from our redshift measurements).

\subsection{Constraints on $t\sub{Q}$ from the projected distribution}
\label{subsec:rdist}

While the redshift distribution is not sensitive for probing shorter QSO
lifetimes, sufficiently small values of $t\sub{Q}$ will affect the
projected distribution of sources in the survey volume. In particular,
the constant-$t\sub{delay}$ paraboloids for
$t\sub{delay}\lesssim 1$ Myr are well matched to the geometry of our
survey volume (see curves on the left side of Fig. \ref{fig:zcombo}, top).

\begin{figure*}
\includegraphics[scale=0.8]{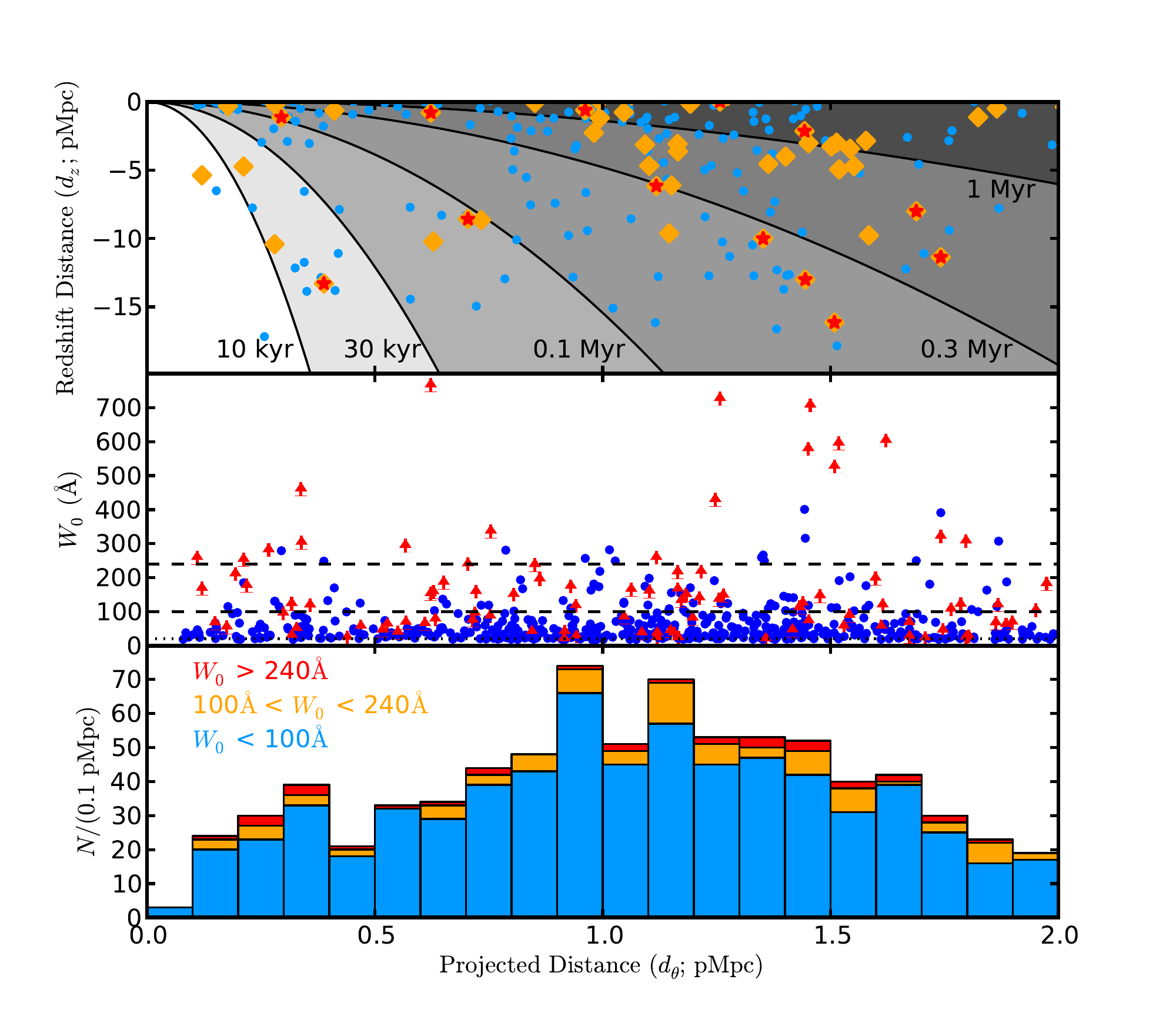}
\caption{Plane-of-sky constraints on QSO lifetimes. {\it Top:}
  Schematic representation of QSO region (as in Fig. \ref{fig:zcombo},
  top), with observer lying to the bottom of the plot. Shaded
  parabolas denote surfaces of constant time delay ($10^4\le
  t\sub{delay}\le 10^6$ as noted on figure). For all lifetimes $t\sub{Q} \lesssim 1$
  Myr, few or no fluorescent sources are predicted at large ($d\sub{$\theta$}\sim 2$
  pMpc) separations from the QSO for our survey geometry. The 2D
  distribution of spectroscopic LAEs is shown for comparison as
  $W_0<100$\AA\ (blue dots), $W_0>100$\AA\ (orange diamonds), and
  $W_0>240$\AA\ (red stars). {\it
    Middle:} \lya\ equivalent width ($W_0$) as a function of
  projected distance from the QSO for photometrically-identified
  LAEs. Red limits and blue points are plotted as in
  Fig. \ref{fig:nbma}. Sources with $W_0>100$\AA\ are common out to
  the maximum projected distances probed by our survey volume,
  suggesting that the QSOs in our sample have lifetimes
  $t\sub{Q}\gtrsim 1$ Myr. {\it Bottom:} The projected
distribution of emitters in three bins of $W_0$, where $W_0$ is either
the detected value or the 1$\sigma$ lower limit. The fraction of
emitters with $W_0>100$\AA\ is fairly constant for $d\sub{$\theta$}<2$
pMpc.}
\label{fig:rcombo}
\end{figure*}

Fig. \ref{fig:rcombo} shows the correspondence of these surfaces
(shown for $10^4$ yr $\le t\sub{delay} \le 10^6$ yr) to
the radial distribution of $W_0$ for the entire photometric sample
(note that the survey geometry is not to scale, unlike in Fig. \ref{fig:zcombo}). As
above, a QSO with $t\sub{Q}\le t\sub{delay}$ will produce
fluorescent emission only within the corresponding paraboloid. For
$t\sub{Q}<0.3$ Myr, this would produce a maximum projected distance
($d\sub{$\theta$}$) at which a fluorescent emitter could appear in our
survey volume. For  $0.3$ Myr $\le t\sub{Q} \le 1$ Myr, the fractional
volume of the survey within the time-delay surface decreases with
increasing $d\sub{$\theta$}$, predicting that the probability of a given
emitter exhibiting fluorescence will likewise decrease with
$d\sub{$\theta$}$. \footnote{The probability of being
  fluorescence-dominated will also vary with distance due to the
  fall-off of the QSO ionizing radiation field, but
  the effect on a flux-limited sample depends on the size
  distribution of fluorescently-emitting regions. Simulations by
  \citet{kol10} indicate that the net effect is small, so we neglect
  it here.}

Notably, the projected distribution of LAEs is well-populated by
objects with $W_0>100$\AA\ and those with $W_0>240$\AA\ out to the
largest values of $d\sub{$\theta$}$ probed by our survey, thus firmly
ruling out lifetimes $t\sub{Q}<0.3$ Myr. Furthermore, the radial
distributions of sources with $W_0>100$\AA\ and $W_0\le 100$\AA\
are entirely consistent when subjected to a two sample KS test
($M\sub{KS}=0.095$; $p<0.24$); the null result is similar using the
$W_0=240$\AA\ threshold. Given the large number of objects in our
photometric sample ($N\sub{phot}=816$), these data provide strong
evidence that $t\sub{Q}\gtrsim 1$ Myr for the QSOs in our sample.

\subsection{Constraints on $\theta\sub{Q}$}
\label{subsec:theta}

For $\theta\sub{Q}\ll \pi/2$, a substantial fraction
of the survey volume at large $d\sub{$\theta$}$ will be inaccessible
to the ionizing emission of the QSO, which will affect the
distribution of $d\sub{$\theta$}$ for fluorescent sources similarly
to $t\sub{Q}\ll1$ Myr. We can probe $\theta\sub{Q}$ in detail through
the 2D spatial distribution of those high-$W_0$ LAEs with spectroscopic
redshifts. It is easily seen in Fig. \ref{fig:rcombo} (top) that the
high-$W_0$ LAEs (red and orange points) extend to large projected
radii at or  near the QSO redshift, suggesting that ionizing emission
is emanating from the QSO nearly perpindicularly to the line of sight
(ie. with $\theta\sub{Q}\sim 90\deg$). In reality, redshift
errors and peculiar velocities prohibit us from establishing the
line-of-sight distance of an LAE from the QSO to be less than 3
pMpc (see \S\ref{subsec:zdist}), so these data (with a 2 pMpc projected
range) provide the constraint  $\theta\sub{Q}\gtrsim
\arctan(2/3)\approx 30\deg$.


\section{Conclusions} \label{sec:conclusions}

We have presented constraints on the lifetime and opening angle of
ionizing QSO emission based on a large photometric/spectroscopic
survey of LAEs in the regions around hyperluminous QSOs (described in
detail in Paper II), finding 1 Myr $\lesssim t\sub{Q} \lesssim$ 20 Myr
and $\theta\sub{Q}\gtrsim 30\degr$.

These results are consistent with the most of the literature
discussed in \S\ref{sec:intro}; in particular, our estimate of
$t\sub{Q}$ falls at the short end of the broad range allowed by the
measurements reviewed in \citet{mar04} and is similar to the
transverse proximity effect
measurements of \citet{gon08}, which included a QSO of 
comparable luminosity to those in this sample. Perhaps most significantly,
$t\sub{Q}$ measured here is short compared to the $e$-folding
timescale of \citet{sal64}:
$t\sub{Sal}=M\sub{BH}/\dot{M}\sub{BH}\approx 45$ Myr
for a QSO with $L=L\sub{Edd}$ and a radiative efficiency
$\epsilon=L/\dot{M}c^2=0.1$. Unless these QSOs have $L\gg L\sub{Edd}$
or are accreting with a low radiative efficiency (neither of
which is expected for luminous QSOs), then these observed hyperluminous
accretion events do not dominate the accretion history of their central
black holes.

%
%


\acknowledgments

\noindent We are indebted to the staff of the W.M. Keck Observatory who keep the
instruments and telescopes running effectively. We also wish to extend
thanks to those of Hawaiian ancestry on whose sacred mountain we are
privileged to be guests. This work has been supported in part by the US
National Science Foundation through grant AST-0908805.





\begin{thebibliography}{28}
\expandafter\ifx\csname natexlab\endcsname\relax\def\natexlab#1{#1}\fi

\bibitem[{{Adelberger} \& {Steidel}(2005)}]{ade05e}
{Adelberger}, K.~L., \& {Steidel}, C.~C. 2005, \apj, 630, 50, 50

\bibitem[{{Adelberger} {et~al.}(2006){Adelberger}, {Steidel}, {Kollmeier}, \&
  {Reddy}}]{ade06}
{Adelberger}, K.~L., {Steidel}, C.~C., {Kollmeier}, J.~A., \& {Reddy}, N.~A.
  2006, \apj, 637, 74, 74

\bibitem[{{Berlind} \& {Weinberg}(2002)}]{ber02}
{Berlind}, A.~A., \& {Weinberg}, D.~H. 2002, \apj, 575, 587, 587

\bibitem[{{Cantalupo} {et~al.}(2012){Cantalupo}, {Lilly}, \&
  {Haehnelt}}]{can12}
{Cantalupo}, S., {Lilly}, S.~J., \& {Haehnelt}, M.~G. 2012, \mnras, 425, 1992,
  1992

\bibitem[{{Cantalupo} {et~al.}(2007){Cantalupo}, {Lilly}, \&
  {Porciani}}]{can07}
{Cantalupo}, S., {Lilly}, S.~J., \& {Porciani}, C. 2007, \apj, 657, 135, 135

\bibitem[{{Cantalupo} {et~al.}(2005){Cantalupo}, {Porciani}, {Lilly}, \&
  {Miniati}}]{can05}
{Cantalupo}, S., {Porciani}, C., {Lilly}, S.~J., \& {Miniati}, F. 2005, \apj,
  628, 61, 61

\bibitem[{{Conroy} \& {White}(2013)}]{con13}
{Conroy}, C., \& {White}, M. 2013, \apj, 762, 70, 70

\bibitem[{{Ferrarese} \& {Merritt}(2000)}]{fer00}
{Ferrarese}, L., \& {Merritt}, D. 2000, \apjl, 539, L9, L9

\bibitem[{{Gebhardt} {et~al.}(2000){Gebhardt}, {Bender}, {Bower}, {Dressler},
  {Faber}, {Filippenko}, {Green}, {Grillmair}, {Ho}, {Kormendy}, {Lauer},
  {Magorrian}, {Pinkney}, {Richstone}, \& {Tremaine}}]{geb00}
{Gebhardt}, K., {Bender}, R., {Bower}, G., {et~al.} 2000, \apjl, 539, L13, L13

\bibitem[{{Gon{\c c}alves} {et~al.}(2008){Gon{\c c}alves}, {Steidel}, \&
  {Pettini}}]{gon08}
{Gon{\c c}alves}, T.~S., {Steidel}, C.~C., \& {Pettini}, M. 2008, \apj, 676,
  816, 816

\bibitem[{{Gould} \& {Weinberg}(1996)}]{gou96}
{Gould}, A., \& {Weinberg}, D.~H. 1996, \apj, 468, 462, 462

\bibitem[{{Hennawi} \& {Prochaska}(2013)}]{hen13}
{Hennawi}, J.~F., \& {Prochaska}, J.~X. 2013, \apj, 766, 58, 58

\bibitem[{{Hogan} \& {Weymann}(1987)}]{hog87}
{Hogan}, C.~J., \& {Weymann}, R.~J. 1987, \mnras, 225, 1P, 1P

\bibitem[{{Jakobsen} {et~al.}(2003){Jakobsen}, {Jansen}, {Wagner}, \&
  {Reimers}}]{jak03}
{Jakobsen}, P., {Jansen}, R.~A., {Wagner}, S., \& {Reimers}, D. 2003, \aap,
  397, 891, 891

\bibitem[{{Kelly} {et~al.}(2010){Kelly}, {Vestergaard}, {Fan}, {Hopkins},
  {Hernquist}, \& {Siemiginowska}}]{kel10}
{Kelly}, B.~C., {Vestergaard}, M., {Fan}, X., {et~al.} 2010, \apj, 719, 1315,
  1315

\bibitem[{{Kollmeier} {et~al.}(2010){Kollmeier}, {Zheng}, {Dav{\'e}}, {Gould},
  {Katz}, {Miralda-Escud{\'e}}, \& {Weinberg}}]{kol10}
{Kollmeier}, J.~A., {Zheng}, Z., {Dav{\'e}}, R., {et~al.} 2010, \apj, 708,
  1048, 1048

\bibitem[{{Kornei} {et~al.}(2010){Kornei}, {Shapley}, {Erb}, {Steidel},
  {Reddy}, {Pettini}, \& {Bogosavljevi{\'c}}}]{kor10}
{Kornei}, K.~A., {Shapley}, A.~E., {Erb}, D.~K., {et~al.} 2010, \apj, 711, 693,
  693

\bibitem[{{Magorrian} {et~al.}(1998){Magorrian}, {Tremaine}, {Richstone},
  {Bender}, {Bower}, {Dressler}, {Faber}, {Gebhardt}, {Green}, {Grillmair},
  {Kormendy}, \& {Lauer}}]{mag98}
{Magorrian}, J., {Tremaine}, S., {Richstone}, D., {et~al.} 1998, \aj, 115,
  2285, 2285

\bibitem[{{Martini}(2004)}]{mar04}
{Martini}, P. 2004, Coevolution of Black Holes and Galaxies, 169, 169

\bibitem[{{Press} \& {Schechter}(1974)}]{pre74}
{Press}, W.~H., \& {Schechter}, P. 1974, \apj, 187, 425, 425

\bibitem[{{Rudie} {et~al.}(2012){Rudie}, {Steidel}, {Trainor}, {Rakic},
  {Bogosavljevi{\'c}}, {Pettini}, {Reddy}, {Shapley}, {Erb}, \& {Law}}]{rud12a}
{Rudie}, G.~C., {Steidel}, C.~C., {Trainor}, R.~F., {et~al.} 2012, \apj, 750,
  67, 67

\bibitem[{{Salpeter}(1964)}]{sal64}
{Salpeter}, E.~E. 1964, \apj, 140, 796, 796

\bibitem[{{Schaerer}(2002)}]{sch02}
{Schaerer}, D. 2002, \aap, 382, 28, 28

\bibitem[{{Seljak}(2000)}]{sel00}
{Seljak}, U. 2000, \mnras, 318, 203, 203

\bibitem[{{Steidel} {et~al.}(2011){Steidel}, {Bogosavljevi{\'c}}, {Shapley},
  {Kollmeier}, {Reddy}, {Erb}, \& {Pettini}}]{ste11}
{Steidel}, C.~C., {Bogosavljevi{\'c}}, M., {Shapley}, A.~E., {et~al.} 2011,
  \apj, 736, 160, 160

\bibitem[{{Trainor} \& {Steidel}(2012)}]{tra12}
{Trainor}, R.~F., \& {Steidel}, C.~C. 2012, \apj, 752, 39, 39

\bibitem[{{Worseck} {et~al.}(2007){Worseck}, {Fechner}, {Wisotzki}, \&
  {Dall'Aglio}}]{wor07}
{Worseck}, G., {Fechner}, C., {Wisotzki}, L., \& {Dall'Aglio}, A. 2007, \aap,
  473, 805, 805

\bibitem[{{Zehavi} {et~al.}(2004){Zehavi}, {Weinberg}, {Zheng}, {Berlind},
  {Frieman}, {Scoccimarro}, {Sheth}, {Blanton}, {Tegmark}, {Mo}, {Bahcall},
  {Brinkmann}, {Burles}, {Csabai}, {Fukugita}, {Gunn}, {Lamb}, {Loveday},
  {Lupton}, {Meiksin}, {Munn}, {Nichol}, {Schlegel}, {Schneider}, {SubbaRao},
  {Szalay}, {Uomoto}, \& {York}}]{zeh04}
{Zehavi}, I., {Weinberg}, D.~H., {Zheng}, Z., {et~al.} 2004, \apj, 608, 16, 16

\end{thebibliography}

%

\clearpage



\begin{deluxetable*}{lccccc}
\tablecaption{Field Descriptions}
\tablewidth{0pt}
\tablehead{
\colhead{QSO Field} & $z\sub{Q}$ & NB Filter & $N\sub{phot}$ &
$N\sub{spec}$ & $N\sub{$W_0$$>$100\AA}$
}

\startdata
Q0100+13 (PHL957) & $ 2.721\pm0.001 $ &   NB4535    & 79  & 20    & 9  \\
HS0105+1619         &  $ 2.652\pm0.001 $ &   NB4430 & 109 & 23     & 6    \\
Q0142$-$10 (UM673a) &  $ 2.743\pm0.001 $ &   NB4535 & 58  & 25    & 11  \\
Q1009+29 (CSO 38) &  $ 2.652\pm0.001 $ &   NB4430   & 60  & 35    & 13  \\
HS1442+2931       &  $ 2.660\pm0.001 $ &   NB4430   & 120 & 39   & 23   \\
HS1549+1919       &  $ 2.843\pm0.001 $ &   NB4670   & 202 & 95 & 27    \\
HS1700+6416         &  $ 2.751\pm0.001 $ &   NB4535 & 66  & 23    & 11  \\
Q2343+12            & $ 2.573\pm0.001 $ &   NB4325  & 122 & 0\tablenotemark{a}     & 16  

\enddata
\tablenotetext{a}{Spectroscopic follow-up observations of field
  Q2343+12 have not yet been obtained.}
\label{table:fields}
\end{deluxetable*}






\end{document}